\documentclass[twocolumn]{aa}

\usepackage{amsmath}
\usepackage{amssymb}
\usepackage{lipsum}
\usepackage{graphics}
\usepackage[varg]{txfonts}
\usepackage{graphicx}
\usepackage{epsfig}
\usepackage{wrapfig}
\usepackage{rotating}
\usepackage{color}
\usepackage{comment}
\usepackage[normalem]{ulem}
\usepackage{multirow}
\usepackage{tabularx}
\usepackage{graphicx}
\usepackage{dcolumn}
\usepackage{hyperref}
\hypersetup{colorlinks=true, allcolors=blue}
\usepackage{float}
\usepackage{natbib}

\usepackage{pgfplots}
\usetikzlibrary{pgfplots.groupplots}
\usepackage{tikz}
\usetikzlibrary{positioning}
\usetikzlibrary{calc}
\usepackage{fp}
\usetikzlibrary{fixedpointarithmetic}

\usepackage{subfig}
\newlength\figureheight 
\newlength\figurewidth 
\usepackage{color}

\begin{document}

\title{A simplified approach for reproducing fully relativistic spectra in X-ray binary systems: Application to Cygnus X-1}

\author{Th. V. Papavasileiou\inst{1, 2}
\and O. Kosmas\inst{1}
\and T. S. Kosmas\inst{1}}
\institute{Department of Physics, University of Ioannina, GR-45110 Ioannina, Greece \\
\email{odykosm@gmail.com, hkosmas@uoi.gr}
\and Department of Informatics, University of Western Macedonia, GR-52100 Kastoria, Greece \\
\email{th.papavasileiou@uowm.gr}}

\date{Received <date> /
Accepted <date>}
  
\abstract
{{Context.} General relativistic effects are strong near the black hole of an X-ray binary and significantly impact the total energy released at the innermost accretion disk's region. The simple pseudo-Newtonian solution in the standard disk model cannot replenish for effects, such as light-bending, gravitational redshift, Doppler boost, etc. Those heavily affect the observed spectra depending on the rotation of the black hole and the disk's inclination.        
 
{Aims.} A proper relativistic treatment would be troublesome, unappealing, or require more advanced computational tools (e.g. the \texttt{kerrbb} code). Our goal is to fully incorporate the black hole's spin and all the general relativistic effects on the observed spectra coming from X-ray binary systems while maintaining the simplicity of the standard Shakura-Sunyaev disk model.         

{Methods.} We propose a way to replicate general relativistic spectra as predicted by the Novikov-Thorne model and the \texttt{kerrbb} numerical code by assuming a standard accretion disk with a shifted inner boundary that depends on the black hole spin and the source's viewing angle. An essential aspect in employing this approach for a broader range of disk inclinations is the derivation of spin-dependent generalized temperature profiles for the accretion disk, obtained from some of the most efficient pseudo-Newtonian potentials around Kerr black holes. We then apply this method to Cygnus X-1, fitting the observational data obtained during its high/soft and hard/low spectral states. 

{Results.} The fully relativistic spectra are reproduced to an excellent approximation with an error margin of 0.03-4\% by a standard disk model with a modified innermost radius within the range of $R_{in}=(0.2-2)R_{ISCO}$, depending on the source's viewing angle and black hole spin. This approach produces observed spectra as predicted by general relativity without the need for the ray-tracing method and complex numerical calculations. Thus, it emerges as a more straightforward alternative way of estimating black hole spins through continuum-fitting by successfully blending the general relativity properties with the Newtonian simplicity in a more complete way than the pseudo-Newtonian solutions. Relativistic effects near the black hole make an otherwise standard accretion disk with inclination $\theta <60^{\circ}$ seem truncated to larger radii to a distant observer. On the other hand, an edge-on view of the disk gives the perspective of being pulled closer to the central object than the respective ISCO radius. In addition, we show that the observational data of Cygnus X-1 can be satisfactorily fitted by employing a reasonably simple lepto-hadronic jet model and a hybrid thermal/non-thermal corona along with the relativistic-equivalent standard thin accretion disk.}


\titlerunning{short title}
\authorrunning{name(s) of author(s)}

\maketitle

\section{Introduction}

Black hole X-ray binaries (BHXRBs) feature an accretion disk of gas and matter extracted from the stellar companion and twin relativistic jets propagating in the interstellar medium. Those systems constitute constant and prominent sources of detectable multi-wavelength emission reaching the Earth, which is measured by prominent space telescopes, such as INTEGRAL, C.T.A., MAGIC, H.E.S.S, and Fermi-LAT \citep{Albert2007, Bodaghee2013, Ahnen2017, Ahnen2018, Malizia_2023}.  

Many studies focus on modeling the broadband emission from X-ray binary systems (XRBs), such as Cygnus X-1 and SS 433. Those studies provide new insights into the jet collimation and acceleration mechanisms, the accretion process, the transition between different spectral states, and the disk-corona dynamics \citep{Gierlinski, Fender2004, Bosch-Ramon2006, Romero2007, Zdziarski_2012, Zhang_2014, Kantzas, Papavasileiou2021, Mastichiadis2022, Papavasileiou_AA, Papavasileiou_2023}. In particular, fitting the accretion disk spectra can provide information about the black hole itself, such as the mass and spin, and how it evolves with time \citep{Malkan_1989, Ghisellini_2009, Calderone_2013, Capellupo_2016}. 

Adapting pseudo-Newtonian (PN) potentials in the standard disk model is a simple way to replenish some of the emission output emerging due to the geometric properties of general relativity \citep{Yang_1995, Hawley_2001, Sarkar_2016, Dihingia_2017}. However, this plain solution is not free of a substantial error margin, especially for rapidly rotating black holes \citep{Artemova_1996, Semerak_1999, Mukhopadhyay_2002, Chakrabarti_2006, Dihingia_2018}. 

Pseudo-potentials also lack the necessary information about how the viewing angle affects the observed spectra due to effects such as gravitational redshift or bending of light near the event horizon. The latter causes a partial reheating of the accretion disk or a decrease in the angular momentum of the compact object due to the returning radiation from the disk \citep{Novikov_1973, Li_2005}. 

The numerical code \texttt{kerrbb} integrates all the relativistic effects according to the Novikov-Thorne thin disk model \citep{Novikov_1973}, including the disk's self-irradiation caused by light deflection \citep{Li_2005}. It is employed as part of the \texttt{xspec} package \citep{Arnaud_1996} for fitting accretion disk spectra through a ray-tracing method providing a black-body emission continuum. Nonetheless, finding a straightforward and simpler way to estimate the fully relativistic spectra from a standard accretion disk is necessary for constructing relativistic-equivalent models to the Shakura-Sunyaev prototype.  

In this work, we aim to reproduce the fully relativistic disk spectra and their dependence on the black hole spin and disk inclination by appropriately modifying the disk's innermost boundary. The latter is achieved based on the emission pattern, as determined by the observed radiative efficiency of Kerr black holes following the \texttt{kerrbb} model \citep{Campitiello_2018}. This approach requires implementing efficient pseudo-Newtonian (PN) gravitational potentials that are applicable even at smaller radii than the Innermost Stable Circular Orbit (i.e., ISCO) limit. Hence, we need to derive valid temperature profiles that depend on the black hole's angular momentum. 

Moreover, this process enables us to avoid the complexities of a complete general relativistic treatment akin to that of \citet{Novikov_1973}. Additionally, it solidifies a method for fully integrating general relativity in fitting the observed spectra from XRBs while upholding the fundamental assumptions and simplicity of the Shakura-Sunyaev model. 

We adopt this method to fit the spectra from Cygnus X-1 in both its hard/low and high/soft spectral states. Towards this aim, we employ the recently revised values for the distance and the mass of the black hole associated with Cygnus X-1 \citep{MillerJones2021}. We also consider an angle to the line of sight that exceeds 45$^{\circ}$ based on recent findings on the system's high polarization degree \citep{Krawczynski_2022}.           

Cygnus X-1 is one of the most well-studied and observed XRBs in our Galaxy. Measurements of its spectral distribution in the soft state by RXTE, ASCA, and OSSE in 1996 \citep{Ling_1997, Dotani_1997} and more recently by XMM-Newton, NuSTAR, NICER, and IXPE \citep{Duro_2011, Walton_2016, Konig_2024, Steiner_2024} indicate a thermal component from the accretion disk and a hard tail up to the MeV scale. The latter is assumed to originate from the up-scattering of seed photons from the disk in a thermal/non-thermal corona hybrid.

Admittedly, several observational data of Cygnus X-1 in its hard/low state dictate various non-thermal emission mechanisms associated with the relativistic jet and the corona. Those include the IR photometry from 0.1-0.5 eV by \citet{Persi_1980} and the radio detection by \citet{Mirabel_1996} and \citet{Fender_2000}. Soft and hard X-ray emission was detected by BeppoSAX and INTEGRAL \citep{DiSalvo_2001, Zdziarski_2012} while COMPTEL measurements occupy a soft gamma-ray tail \citep{McConnell_2002}. Higher-energy detection attempts and upper limits are carried out by the Fermi-LAT and MAGIC telescopes \citep{Malyshev_2013, Ahnen2017}.

In the remainder of this work, we discuss the disk geometry and the radial temperature obtained from the two PN solutions referring to rotating black holes in Sec. \ref{Sec2}. In Sec. \ref{Sec3}, we discuss the radiative efficiency perceived by an observer and how it can be employed to acquire an expression for the location of the disk's inner edge as a function of the spin and the viewing angle. Then, in Sec. \ref{Sec4}, we compare our approach to general relativistic results and apply this method to the multi-wavelength spectra detected from Cygnus X-1 in its two primary spectral states. Finally, we conclude our results in Sec. \ref{Sec5}.    

\section{Spin incorporation in the dynamics of the standard disk model}\label{Sec2}  

\subsection{Marginally stable orbits around rotating black holes}

The stellar black holes in XRBs introduce some fundamental general relativity properties. They perplex the calculations and challenge the validity of the simplistic Newtonian hydrodynamic treatment of the standard model. Those properties are the locations of the marginally bound orbits (i.e., $R_{mb}$) and the marginally stable circular orbits (i.e., $R_{ms}$ or $R_{ISCO}$). The first case refers to particles not succumbing to the gravitational pull beyond the event horizon, while the latter constitutes the innermost boundary of the accretion disk.

Static black holes are characterized by an ISCO (i.e., Innermost Stable Circular Orbit) radius of $6R_g$, with $R_g=GM_{bh}/c^2$ being the gravitational radius. However, most black holes display a rotational motion that leads to the accretion disk's size extension depending on the corresponding spin parameter. Therefore, the disk begins at a radius given by \citep{Bardeen_1972} 
\begin{equation}
\label{isco}
R_{in}= R_{ISCO}= \frac{GM_{bh}}{c^{2}}\left(3+\lambda_{2}\pm\sqrt{(3-\lambda_{1})(3+\lambda_{1}+2\lambda_{2})}\right) ,
\end{equation}  
where it is
\begin{align}
\lambda_{1} &= 1+\left(1-\alpha _{*}^{2}\right)^{1/3}\left[\left(1+\alpha _{*}\right)^{1/3}+\left(1-\alpha _{*}\right)^{1/3}\right] , \\
\lambda_{2} &= \left(\lambda_{1}^{2}+3\alpha _{*}^{2}\right)^{1/2} . 
\end{align} 
The dimensionless spin parameter $\alpha _{*}=cJ/GM_{bh}^{2}$ varies between $0$ and $1$. Moreover, the maximum possible value is considered to be $\alpha _{*}\approx 0.998$, although indications suggest a higher limit considering a geometrically thick accretion flow \citep{Benson_2009}. In Eq. (\ref{isco}), the plus sign corresponds to the disk's retrograde orbit (i.e., opposite rotational direction) around the black hole. On the other hand, the minus sign corresponds to a prograde one (i.e., co-rotation), which is the case for most binary systems.

\begin{figure}[ht] 
\begin{tikzpicture}
\centering
  \node (img1)  {\includegraphics[width=\linewidth]{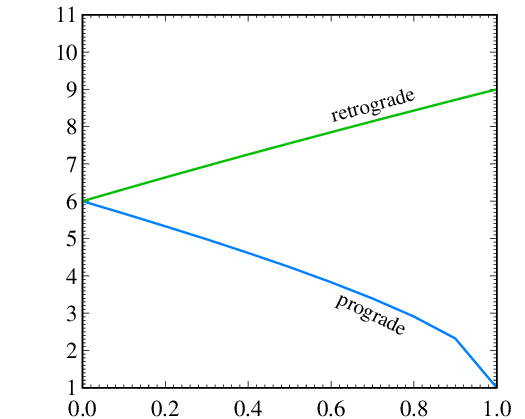}};
  \node[left= of img1, node distance=0cm, rotate=90, anchor=center,yshift=-1.7cm, font=\color{black}, font=\normalsize] {$R_{ISCO}/R_{g}$};
  \node[below= of img1, node distance=0cm, yshift=1cm, xshift=0.5cm, font=\color{black}, font=\normalsize] {$\alpha _{*}$};
\end{tikzpicture}

\caption{\label{figure1} The Innermost Stable Circular Orbit (ISCO) radius as a function of the black hole's spin corresponding to prograde and retrograde orbits of the accretion flow.}
\end{figure}

The ISCO radius varies between $R_{g}$ and $9R_{g}$ depending on the rotational velocity and direction. Regarding prograde orbits, a significant increase of $\alpha _{*}$ leads to a disk augmentation up to the gravitational radius, as seen in Fig. \ref{figure1}. On the other hand, maximum counter-rotation leads to a truncated accretion disk with a maximized inner boundary at $9R_{g}$.      

\subsection{The pseudo-Newtonian solution to the standard accretion disk}

Regarding static black holes, a pretty successful treatment employs the PN potential proposed by Paczyński and Wiita \citep{Paczynsky_1980, Abramowicz_2009}. Calculating the disk's angular velocity using the following potential $\Phi_{PW}= -GM_{bh}/(R-2R_g)$ yields the fixed disk temperature profile, assuming a torque-free boundary condition \citep{Gierlinski}. This condition is a pretty usual assumption, which is even more valid for rotating central objects since non-zero torque effects are less significant. Moreover, this treatment is often employed for rotating black holes, neglecting its limitations regarding the frame-dragging caused by the Lense-Thirring effect \citep{Pfister_2007}. 

Accounting for the black hole's angular momentum in the standard model requires a new, convenient solution in the same manner as the Paczyński-Wiita, introducing a generalized radial temperature. One of the most well-known and successful Kerr pseudo-potentials is the one derived by \citet{Artemova_1996}. The corresponding gravitational force is given by
\begin{equation}
\label{A96}
F_A= \frac{GM_{bh}}{R_{g}^2}\frac{1}{r^{2-\beta}(r-r_h)^{\beta}} , 
\end{equation}
where $r=R/R_g$, $r_h= 1+\sqrt{1-\alpha _{*}^2}$ is the event horizon in the Kerr metric, and $\beta = r_{in}/r_h-1$ denotes a dimensionless constant for a given specific black hole angular momentum. The corresponding potential reproduces pretty accurately the location of the marginally stable orbit and the disk radiative efficiency as predicted by general relativity.

Another pseudo-Newtonian solution derived directly from the Kerr metric was proposed by \citet{Mukhopadhyay_2002} as
\begin{equation}
\label{M02}
F_M= \frac{GM_{bh}}{R_{g}^2}\frac{\left(r^2-2\alpha _{*}\sqrt{r}+\alpha _{*}^2\right)^2}{r^3\left[\sqrt{r}(r-2)+\alpha _{*}\right]^2} .
\end{equation}
This expression refers to the equatorial plane and reproduces $R_{mb}$ within 5\% of error and the mechanical energy at $R_{ISCO}$ with a 10\% maximum deviation from the theoretical predictions. Hence, it produces a slightly increased luminosity in comparison. The energy dissipation distribution exhibits a less than 10\% error for extreme spin values. Negative values of $\alpha _{*}$ correspond to a counter-rotating accretion flow relative to the black hole. Both Eqs. (\ref{A96}) and (\ref{M02}) reduce to the Paczyński-Wiita formula for $\alpha _{*}=0$. 

The Mukhopadhyay solution has limited use due to its complex and troublesome analytical expression of the potential. However, we can still derive valid temperature profiles for the accretion disk using the force expression. To confirm our results, we use numerical integration of the gravitational force.

More recently, \citet{Dihingia_2018} improved upon the attempt of \citet{Chakrabarti_2006} to study accretion flows around rotating black holes with $\alpha _{*}<0.8$. They nullify the spin limitation since most observational data incline towards extreme rotating stellar objects. However, the proposed effective potential is not as convenient or accurate in its predictions compared to the ones discussed above.   

\subsection{Spin-dependent disk temperature profiles}

The radial temperature of an accretion disk around a Kerr black hole employing the pseudo-potential of Eq. (\ref{A96}) is derived as 
\begin{equation}
\label{temp_A}
T_A(r)= T_0\left(\frac{(3-\beta )(r-r_h)+\beta r}{3r^{3-\beta}(r-r_h)^{\beta +1}}\right)^{1/4}\left(1-\frac{G_A(r_{in})}{G_A(r)}\right)^{1/4} ,   
\end{equation}
where we have
\begin{equation}
G_A(r)= \left(\frac{r^{\beta +1}}{(r-r_h)^{\beta}}\right)^{1/2} .
\end{equation}
The constant $T_0$ depends on the system's accretion rate and the black hole's mass as
\begin{equation}
T_0= \left(\frac{3GM_{bh}\dot{M}}{8\pi\sigma R_{g}^3}\right)^{1/4} .
\end{equation}
The mathematical expressions used to derive $T(r)$ can be found in Appendix \ref{Temp_deriv}. In a similar way, the modified temperature profile that corresponds to the Mukhopadhyay potential is written as 
\begin{equation}
\label{temp_M}
T_M(r)= T_0\left(\frac{4G_M(r)^2}{3r^4}-W(r)\right)^{1/4}\left(1-\frac{G_M(r_{in})}{G_M(r)}\right)^{1/4} .   
\end{equation}
The spin contribution is inserted through the following functions
\begin{align}
G_M(r)&= \frac{r^2-2\alpha _{*}\sqrt{r}+\alpha _{*}^2}{\sqrt{r}(r-2)+\alpha _{*}} , \\
W(r)&= \frac{2G_M(r)}{3r^3\left(\sqrt{r}(r-2)+\alpha _{*}\right)}\left(2r-\frac{\alpha _{*}}{\sqrt{r}}-\frac{3r-2}{2\sqrt{r}}G_M(r)\right) .
\end{align}
In comparison, Eq. (\ref{temp_A}) produces negligibly larger disk luminosity across the spectrum for energies below 1 keV that drops slightly more abruptly for higher energies. The deviation is marginal for large $\alpha _{*}$ in the co-rotation case, while they coincide for the rest of the spin values. Both expressions reduce to the temperature function of \citet{Gierlinski} that implements the Paczyński-Wiita potential. 

\begin{figure}[ht] 
\begin{tikzpicture}
\centering
  \node (img1)  {\includegraphics[width=\linewidth]{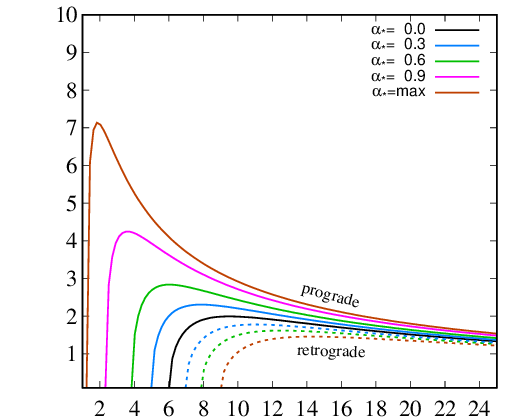}};
  \node[left= of img1, node distance=0cm, rotate=90, anchor=center,yshift=-1.4cm, font=\color{black}, font=\normalsize] {$T$ [$\times 10^6 K$]};
  \node[below= of img1, node distance=0cm, yshift=1cm, xshift=0.5cm, font=\color{black}, font=\normalsize] {$r/r_g$};
\end{tikzpicture}

\caption{\label{figure2} Temperature dependence on the disk's radius for different values of the black hole spin.}
\end{figure}

Fig. \ref{figure2} demonstrates the disk surface temperature of Eq. (\ref{temp_M}) for varying positive and negative spins, respectively.  As described in the previous subsection, the inner disk edge $r_{in}$ (in units of $R_g$) is given by the assumption of locally non-rotating frames in the Kerr geometry. We can see a significant increase in temperature for $\alpha _{*}>0.9$. It should be noted that this is the value range in which the PN solution is more likely to deviate from a thorough general relativistic treatment.   

The total energy radiated from the accretion disk equals the loss of mechanical energy at its innermost boundary 
\begin{equation}
L_{disk}= -\dot{M}e_{in} ,
\end{equation}
where the specific energy is defined as
\begin{equation}
e= \Phi +\frac{R}{2}\frac{d\Phi}{dR} .
\end{equation}

\begin{figure}[ht] 
\begin{tikzpicture}
\centering
  \node (img1)  {\includegraphics[width=\linewidth]{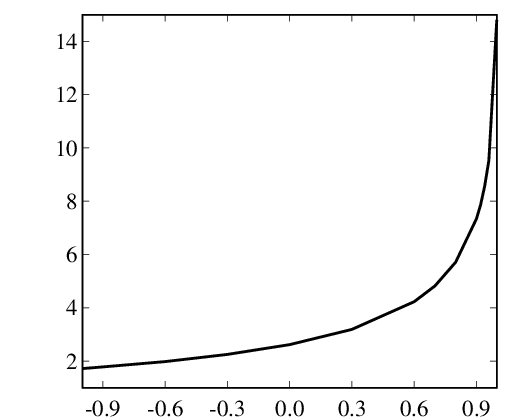}};
  \node[left= of img1, node distance=0cm, rotate=90, anchor=center,yshift=-1.4cm, font=\color{black}, font=\normalsize] {$L_{disk}$ [$\times 10^{38} erg/s$]};
  \node[below= of img1, node distance=0cm, yshift=1cm, xshift=0.5cm, font=\color{black}, font=\normalsize] {$\alpha _{*}$};
\end{tikzpicture}

\caption{\label{figure3} The compact object rotation's impact on the total accretion disk luminosity.}
\end{figure}

In Fig. \ref{figure3}, we can clearly see how the brightness of the disk is determined by the rotation of the black hole. The abrupt increase with higher spin is attributed to the accretion flow near the gravitational radius.

Extreme Kerr black holes (i.e., $\alpha _{*}>0.9$), governed by frame-dragging effects near their event horizon, are verified as central objects for an increasing number of well-known XRBs \citep{McClintock_2006, Stuchlik_2016, Zhao_2021}. Therefore, the spin-dependent profiles given by Eqs. (\ref{temp_A}) and (\ref{temp_M}) yield a more realistic measurement of the energy released near those sources.         

\section{Spectral angular distribution via disk surface fluctuations}\label{Sec3}

The detectable spectra from accretion disks are heavily dependent on the observer's viewing angle. Ignoring all relativistic effects, the observed radiative efficiency is $\eta _{rad}^{obs}=2cos\theta\eta _{rad}(R_{ISCO})$, where $\eta _{rad}=L_{disk}/\dot{M}c^2$. Regardless, frame-dragging, gravitational redshift, Doppler boost, and light-bending effects significantly alter the angular dependence of the perceived emission flows near the black hole. Hence, the emerging emission pattern following the relativistic \texttt{kerrbb} model is determined by an analytic expression derived by \citet{Campitiello_2018} as $\eta _{rad}^{obs}=f(\theta , \alpha _*)\eta _{rad}(R_{ISCO})$. The function $f(\theta , \alpha _*)$ is given in the Appendix \ref{f_func}.

\begin{figure}[ht] 
\begin{tikzpicture}
\centering
  \node (img1)  {\includegraphics[width=\linewidth]{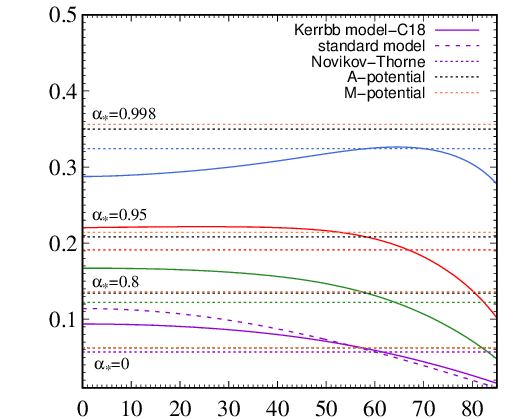}};
  \node[left= of img1, node distance=0cm, rotate=90, anchor=center,yshift=-1.7cm, font=\color{black}, font=\normalsize] {$\eta _{rad}$};
  \node[below= of img1, node distance=0cm, yshift=1cm, xshift=0.5cm, font=\color{black}, font=\normalsize] {$\theta$ [$^{\circ}$]};
\end{tikzpicture}

\caption{\label{figure4} The solid lines correspond to the radiative efficiency given by the analytic counterpart of the \texttt{kerrbb} model as observed for different system inclinations and black hole spin values (referring to \citet{Campitiello_2018} as C18). The purple dashed line showcases the angular dependence corresponding to a Shakura-Sunyaev disk around a static black hole, neglecting all relativistic contributions. The colored dotted lines for each spin value depict the respective theoretical predictions of \citet{Novikov_1973}. For comparison, the respective black and orange-dotted lines represent the Kerr pseudo-Newtonian treatment of this paper.}
\end{figure}

The corresponding observed total luminosity normalized by $\dot{M}c^2$ is plotted for varying values of $\alpha _{*}$ in Fig. \ref{figure4}. We also include the radiative efficiency produced with both pseudo-potentials discussed in this work and the respective theoretical predictions of Novikov-Thorne. The comparison between the PN adaptation and general relativity reveals an increasing deviation for highly rotating black holes.

Concerning non-rotating central objects, the angular distribution, according to the standard model, follows a similar pattern to that of the relativistic case. Higher angular momentum yields an abrupt drop in the total emission output for accretion disks inclined beyond 60$^{\circ}$, a viewing angle less affected by light-bending effects. However, extreme rotation triggers a resurgence between 50$^{\circ}$-70$^{\circ}$ in the radiation escaping from the black hole's gravitational field.       

\subsection{Modification of the disk's inner boundary}

\begin{figure}[ht] 
\begin{tikzpicture}
\centering
  \node (img1)  {\includegraphics[width=\linewidth]{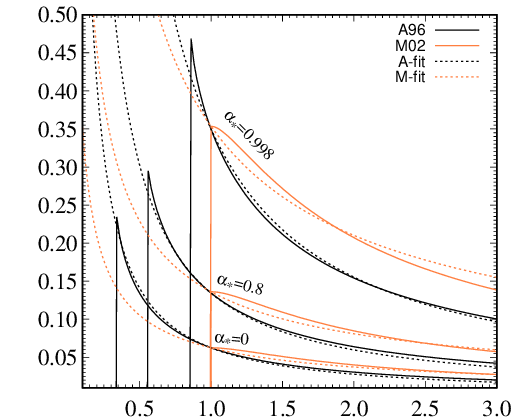}};
  \node[left= of img1, node distance=0cm, rotate=90, anchor=center, yshift=-1.5cm, font=\color{black}, font=\normalsize] {$\eta _{rad}$};
  \node[below= of img1, node distance=0cm, yshift=1cm, xshift=0.5cm, font=\color{black}, font=\normalsize] {$R_{in}/R_{ISCO}$};
\end{tikzpicture}

\caption{\label{figure5} The radiative efficiency for different $\alpha _*$ as a function of the disk's inner radius, given in terms of the $R_{ISCO}$. The total disk luminosity is calculated using the Artemova (as A96, black solid lines) and the Mukhopadhyay (as M02, orange solid lines) potentials. The black dotted lines correspond to Eq. (\ref{eta_Rin}), while the orange ones fit a similar expression with a power-law index of $p=0.75$.}
\end{figure}

The radiative efficiency is independent of the accretion rate and the spectral hardening factor $f_{col}$. It depends negligibly on the black hole's stellar mass $M_{bh}$ and the disk's inner radius $R_{in}$, as long as the outer radius is large enough, $R_{out}>10^{11} \ cm$, akin to the default value set in \texttt{kerrbb}. Fig. \ref{figure5} illustrates this correlation, which is established by employing Eqs. (\ref{temp_A}) and (\ref{temp_M}) in calculating the thermal spectra from the disk. In this work, we fit the power-law form of the Artemova solution (black dotted lines in Fig. \ref{figure5}) with the following expression
\begin{equation}
\label{eta_Rin}
\eta _{rad}(R_{in})= \eta _{rad}(R_{ISCO})\left(\frac{R_{in}}{R_{ISCO}}\right)^{-1.17} .
\end{equation}   
The respective orange solid lines of the Mukhopadhyay potential are fitted using a power-law index of $p=0.75$ instead. A noticeable deviation for all spin values, especially near the ISCO radius, deter us from further exploiting the Mukhopadhyay pseudo-force. Also, the fact that the Artemova potential works for radii significantly smaller than $R_{ISCO}$ makes it a very suitable tool for matching the fully relativistic spectrum for a given angle of view.  

\begin{figure}[ht] 
\begin{tikzpicture}
\centering
  \node (img1)  {\includegraphics[width=\linewidth]{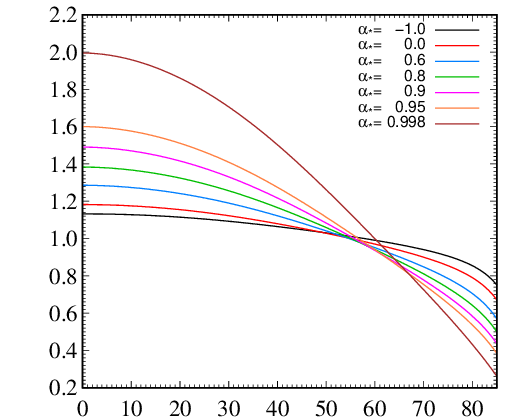}};
  \node[left= of img1, node distance=0cm, rotate=90, anchor=center, yshift=-1.5cm, font=\color{black}, font=\normalsize] {$R_{in}/R_{ISCO}$};
  \node[below= of img1, node distance=0cm, yshift=1cm, xshift=0.5cm, font=\color{black}, font=\normalsize] {$\theta$ [$^{\circ}$]};
\end{tikzpicture}

\caption{\label{figure6} Depending on the source's viewing angle and the black hole's rotation, the shifted innermost edge of the accretion disk yields spectra similar to those obtained with the \texttt{kerrbb} model.}
\end{figure}

Furthermore, including all the effects of general relativity and incorporating the correct angular distribution urges the equivalence between Eq. (\ref{eta_Rin}) and the observed efficiency given by the analytic function of \citet{Campitiello_2018} (see Appendix \ref{f_func}). By doing so, we find an expression for the modified inner boundary of the accretion disk, which is given by
\begin{equation}
\label{R_in}
R_{in}= \left(\frac{2cos\theta}{f(\theta , \alpha _*)}\right)^{1/1.17}R_{ISCO} .
\end{equation} 
This way, equivalent energy distributions to the fully relativistic spectra are reproduced by a standard disk model with the temperature profile of Eq. (\ref{temp_A}) and an innermost radius given by Eq. (\ref{R_in}). The latter is plotted in Fig. \ref{figure6} and depends on the inclination and angular momentum of the accreting compact object. Basically, when we modify the disk inner radius in order to achieve the \texttt{kerrbb} observed radiative efficiency, we produce spectra that depend on the angle $\theta$ in the same way as dictated by general relativity. 

Fig. \ref{figure6} shows that, independently of the rotation, the physical processes associated with an accretion disk observed from an angle of approximately $\theta =60^{\circ}-65^{\circ}$ identify with the Newtonian-like dynamics of the Shakura-Sunyaev model. Thus, a minimum shift of $R_{in}$ is required to fit the corresponding \texttt{kerrbb} spectra. Consequently, the boundary-modification approach is exceptionally accurate regarding this angle range. 

Relativistic effects near the black hole make an otherwise standard accretion disk viewed with an angle of $\theta <60^{\circ}$ seem truncated to larger radii to a distant observer. On the other hand, an edge-on view of the disk gives the perspective of being pulled closer to the central object than the respective ISCO radius. That is a focal point in fabricating fully relativistic counterparts to simple standard disks using Eq. (\ref{R_in}).          

\section{Results and discussion}\label{Sec4}     

\subsection{Comparison to General Relativity}

There are three primary sources of error concerning the approach presented in this work for calculating relativistic spectra. The first is the intrinsic error margin of the analytic function $f(\theta ,\alpha _*)$, which is relatively small and does not exceed 1\% \citep{Campitiello_2018}. The second one is due to the PN-potential adoption, which can reach 8-9\% in the case of the Artemova solution. Lastly, the error margin introduced in the context of this work is due to the deviation between the PN-generated $\eta _{rad}(R_{in})$ and Eq. (\ref{eta_Rin}), with the choice of the $p=1.17$ index, which is also evident in Fig. \ref{figure5}.   

\begin{figure}[ht] 
\begin{tikzpicture}
\centering
  \node (img1)  {\includegraphics[width=\linewidth]{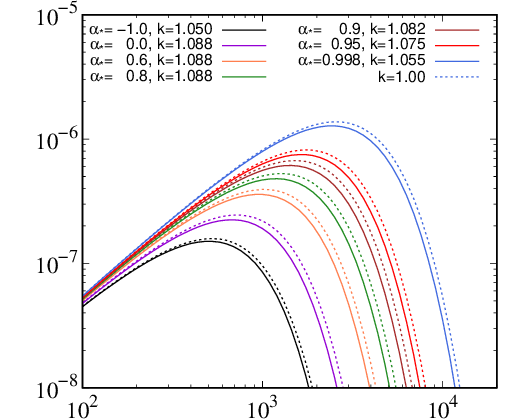}};
  \node[left= of img1, node distance=0cm, rotate=90, anchor=center,yshift=-1.4cm, font=\color{black}, font=\normalsize] {$\epsilon F_{\epsilon}$ [$ergs^{-1}cm^{-2}$]};
  \node[below= of img1, node distance=0cm, yshift=1cm, xshift=0.5cm, font=\color{black}, font=\normalsize] {$E$ [$eV$]};
\end{tikzpicture}

\caption{\label{figure13} Modified disk spectra that correspond to the correct radiative efficiencies for different $\alpha _*$. They are produced by slightly shifting the ISCO radius as $R'_{ISCO}=kR_{ISCO}$ while the rest of the free parameters remain unchanged. The unaltered spectra are also demonstrated (dotted lines) for comparison.}
\end{figure}

By adjusting the disk’s inner radius, we eradicate the most significant part of the PN-induced error entering the calculations. That holds for every choice of parameterization. The only deviation remaining concerns the radiative efficiency at the ISCO radius, $\eta _{rad}(R_{ISCO})$, which is still slightly overestimated. A fitting solution involves a subtle redefinition of the $R_{ISCO}$ according to the factor $k= R'_{ISCO}/R_{ISCO}$, which on average is $k\approx 1.075$, covering the spin value range. The resulting spectra are slightly shifted to lower energies, as shown in Fig. \ref{figure13} for a sample of spin values with their respective shifting factors. For comparison, we also demonstrate the disk emission (dotted lines) assuming the $R_{ISCO}$ given in Eq. (\ref{isco}).   

\begin{figure}[ht] 
\begin{tikzpicture}
\centering
  \node (img1)  {\includegraphics[width=\linewidth]{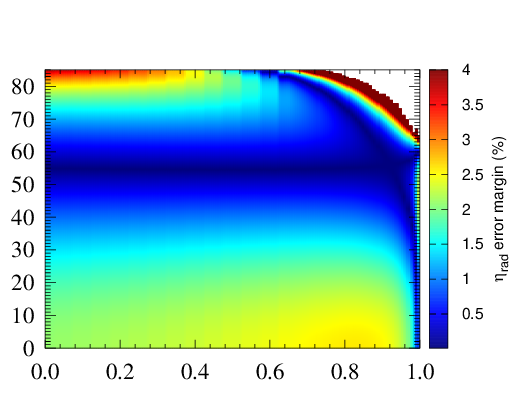}};
  \node[left= of img1, node distance=0cm, rotate=90, anchor=center, yshift=-1.0cm, font=\color{black}, font=\normalsize] {$\theta$ [$^{\circ}$]};
  \node[below= of img1, node distance=0cm, yshift=1.7cm, xshift=-0.4cm, font=\color{black}, font=\normalsize] {$\alpha _*$};
\end{tikzpicture}

\caption{\label{figure14} The $\eta _{rad}$ error heat map depends on the viewing angle and black hole's spin. The observed radiative efficiency is obtained via the boundary-shifting method for approximating the fully relativistic spectra from standard accretion disks.}
\end{figure}

The error introduced due to the index $p=1.17$ in Eq. (\ref{eta_Rin}) ranges between 0.03-2.5\% for $\theta <60^{\circ}$ regardless how fast the black hole is rotating, as demonstrated in the heat map of Fig. \ref{figure14}. Moreover, there is a slight general increase in the deviation for viewing angles close to 80$^{\circ}$ (i.e., \texttt{kerrbb} code operates in the 0-85$^{\circ}$ range) when the spin does not belong in the $\alpha _*=0.4-0.7$ regime. 

Application of this work's approach in highly-inclined accretion disks around co-rotating black holes with $\alpha _*>0.8$ is constrained due to the limitations of the temperature of Eq. (\ref{A96}). That is also obvious from the abruptly broken power-law in Fig. \ref{figure5}. Furthermore, relativistic spectra can be attained for $\theta \le 65^{\circ}$ when $\alpha _*\geq 0.998$. That is somewhat restrictive but falls within the parameter range of many well-known X-ray binaries in our galaxy, while the rest of the sources would require the \texttt{kerrbb} model. 

In Fig. \ref{figure14}, the dark red contour around the restricted white area signifies a substantial increase in the $\eta _{rad}$ error that reaches 10-12\%. Nonetheless, a better-suited index in Eq. (\ref{eta_Rin}) in the $p=1.05-1.2$ range for a specific spin-inclination combination could significantly reduce the deviation from the actual relativistic results. That is one of our targets in future projects.      

\begin{figure*}[ht] 
\begin{tikzpicture}
\centering
  \node (img1)  {\includegraphics[width=0.5\linewidth]{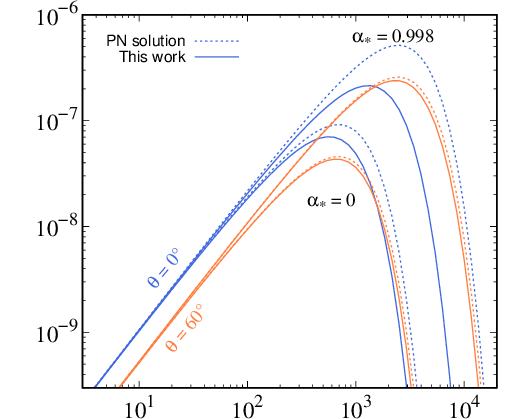}};
  \node[left= of img1, node distance=0cm, rotate=90, anchor=center,yshift=-1.4cm, font=\color{black}, font=\normalsize] {$\epsilon F_{\epsilon}$ [$ergs^{-1}cm^{-2}$]};
  \node[below= of img1, node distance=0cm, yshift=1cm, xshift=0.5cm, font=\color{black}, font=\normalsize] {$E$ [$eV$]};
  \node[right= of img1, xshift=-1.5cm] (img2)  {\includegraphics[width=0.5\linewidth]{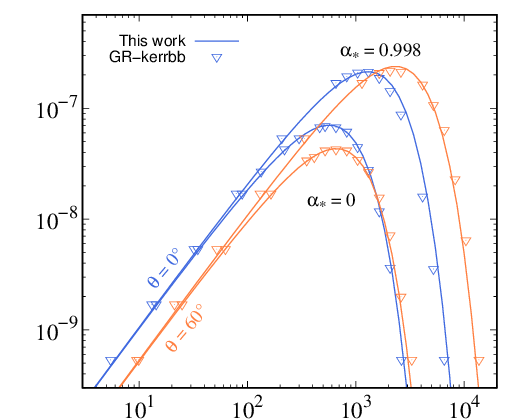}};
  \node[below= of img2, node distance=0cm, yshift=1cm, xshift=0.5cm, font=\color{black}, font=\normalsize] {$E$ [$eV$]};
\end{tikzpicture}

\caption{\label{figure7} Comparison between the disk spectra produced through the boundary-shifting method of this work and (left panel) the Artemova pseudo-Newtonian potential, (right panel) the general relativistic treatment of the \texttt{kerrbb} model. The parameters used are $M_{bh}=10M_{\odot}$, $\dot{M}=10^{18}\ g/s$, and $f_{col}=1$.}
\end{figure*}

In the left panel of Fig. \ref{figure7}, we compare the observed disk spectra corresponding to the PN solution of Eq. (\ref{temp_A}) with the spectra obtained through this work's boundary-shifting approach according to Eq. (\ref{R_in}). We assume a static ($\alpha _*=0$), and a rapidly rotating black hole ($\alpha _*=0.998$) for greater variety. We can see that the two different approaches are rather close for both spin values when the disk is viewed with an angle of $\theta =60^{\circ}$, a result also verified by Fig. \ref{figure6} where $R_{in}\approx R_{ISCO}$. However, the significant deviation is obvious when $\theta =0^{\circ}$, especially in the maximum spin case where the disk's inner edge is closer to the black hole and general relativistic effects are very important. 

The PN potential clearly overestimates the total disk luminosity for $\theta\lesssim 60^{\circ}$, while it underestimates it when $\theta\gtrsim 60^{\circ}$. Besides the potential's intrinsic error, the main reason is that while it offers a very good measurement of the additional energy radiated away by the disk, it cannot offer any information on the spectra a distant observer receives depending on $\theta$. Instead, the PN solution adopts the angular dependence of the classic Shakura-Sunyaev model, $\eta _{rad}^{obs}=2cos\theta\eta _{rad}$. 

Our results show an excellent agreement with the fully relativistic formalism of Novikov and Thorne incorporated in the \texttt{kerrbb} model, as seen in the right panel of Fig. \ref{figure7}. The deviation, partially due to the $\eta _{rad}$ error margin of Fig. \ref{figure14}, is minimal even for highly-inclined accretion disks around rapidly rotating central objects. 

Furthermore, the approach of this work presents a direct alternative solution in calculating the observed fully relativistic disk spectra without any complex mathematical formalism and the need for a ray-tracing method. Given that the Novikov-Thorne model has been proven very efficient by GRMHD simulations \citep{Kulkarni_2011}, our approach could provide a very suitable and straightforward way for estimating black hole spins via the continuum-fitting method.

\begin{table}
\caption{\label{Table1} Cygnus X-1 parameters employed in this work.}
\begin{center}
\begin{tabular}{l l l}
\hline \\ [0.005ex] 
Parameter & Cyg X-1 & Units\\ 
\hline \\ [0.01ex]
$M_{BH}$ & 21.2 & \(M_\odot\) \\ [0.2ex]
$M_{donor}$ & 40.6 & \(M_\odot\) \\ [0.2ex]
$d$ & 2.22 & kpc \\ [0.2ex]
$\theta$ & 60 & $^{\circ}$ \\ [0.2ex]
$P_{orb}$ & 5.5998 & days \\ [0.2ex]
$\alpha _{*}$ & 0.97 & - \\ [0.2ex]
$L_{star}$ & $3.95\times 10^{5}$ & \(L_\odot\) \\ [0.2ex]
$T_{star}$ & 31138 & K \\ [0.2ex]
\hline
\end{tabular}
\end{center}
\textbf{References}. \citet{MillerJones2021}, \citet{Brocksopp1999}, \citet{Krawczynski_2022}, \citet{Stirling}, \citet{Gou_2014}, \citet{Walton_2016}, \citet{Tetarenko2019}
\end{table} 

\subsection{Cygnus X-1: high/soft spectral state}

\begin{figure}[ht] 
\begin{tikzpicture}
\centering
  \node (img1)  {\includegraphics[width=\linewidth]{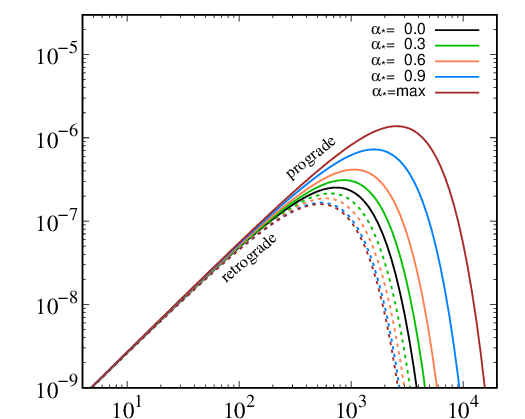}};
  \node[left= of img1, node distance=0cm, rotate=90, anchor=center,yshift=-1.4cm, font=\color{black}, font=\normalsize] {$\epsilon F_{\epsilon}$ [$ergs^{-1}cm^{-2}$]};
  \node[below= of img1, node distance=0cm, yshift=1cm, xshift=0.5cm, font=\color{black}, font=\normalsize] {$E$ [$eV$]};
\end{tikzpicture}

\caption{\label{figure8} The spectral energy distribution (SED) emitted from the accretion disk for varying black hole spins.}
\end{figure}

Using Eqs. (\ref{temp_A}), (\ref{R_in}), and the recently redefined Cygnus X-1 parameters of Table \ref{Table1}, we showcase the spin-dependent disk spectra while the system is going through the soft state in Fig. \ref{figure8}. The solid lines correspond to the alignment between the black hole and accretion disk rotational directions. It should be mentioned that those results were obtained for a fixed mass accretion rate of $\dot{M}=8.5\times 10^{-8}$ \(M_\odot\)/yr, which, in most cases, is a free model parameter. In addition, we estimate the radiative efficiency of an extreme rotating black hole to be $\eta _{rad}\approx 0.35$, which agrees with the theoretical predictions. Moreover, employing this work's approach allows us to constrain more accurately the black hole's accretion rate as long as we have a precise value of the spin or vice versa.  

The spectrum's shift to higher energies in Fig. \ref{figure8} is attributed to a substantially hotter inner disk layer that contributes heavily to the harder X-ray thermal component. On the other hand, the soft thermal emission corresponds to the extended disk region far from the ISCO radius (i.e., $R\gg 6R_g$).

\begin{figure}[ht] 
\begin{tikzpicture}
\centering
  \node (img1)  {\includegraphics[width=\linewidth]{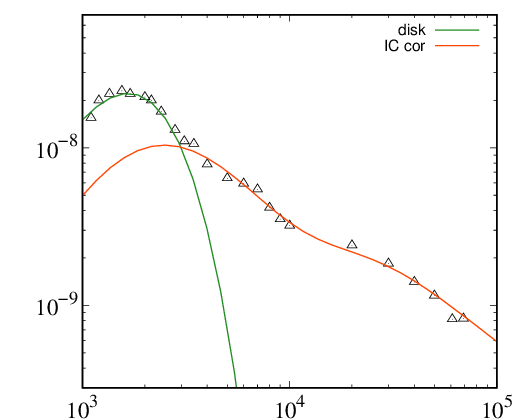}};
  \node[left= of img1, node distance=0cm, rotate=90, anchor=center,yshift=-1.4cm, font=\color{black}, font=\normalsize] {$\epsilon F_{\epsilon}$ [$ergs^{-1}cm^{-2}$]};
  \node[below= of img1, node distance=0cm, yshift=1cm, xshift=0.5cm, font=\color{black}, font=\normalsize] {$E$ [$eV$]};
\end{tikzpicture}

\caption{\label{figure9} The high/soft spectral state of Cygnus X-1 was observed by RXTE, ASCA, and NuSTAR \citep{Ling_1997, Dotani_1997, Walton_2016}. The observational data are fitted by a black-body spectrum emitted by the inner disk region and an IC scattered component of the disk emission off a combination of thermal and non-thermal corona electrons.}
\end{figure}

In Fig. \ref{figure9}, we outline the data obtained during observations of the soft spectral state of Cygnus X-1 \citep{Ling_1997, Dotani_1997, Walton_2016}. The thermal emission at lower energies is fitted by a black-body spectrum emitted by the inner disk regions. The respective spin is found to be $\alpha _{*}=0.97$, satisfying the lower limit $\alpha _*>0.9696$ set by \citet{MillerJones2021} for the slightly misaligned spin-orbit case. This value is close to the limits set by the continuum-fitting method as $\alpha _*>0.983$ \citep{Gou_2014} and the relativistic reflection modeling as $0.93\gtrsim\alpha _*\lesssim 0.96$ \citep{Walton_2016}. The high spin yields an accretion disk close to the compact object (i.e., $R_{in}\approx R_g$), a primary characteristic of the high/soft state. 

Non-blackbody effects are incorporated through a spectral hardening factor $f_{col}\approx 1-2$ in multi-color disk models \citep{Merloni_2000, Li_2005}. It constitutes a color correction of the disk temperature. This adjustment, paired with an appropriate accretion rate, accounts for the intense electron scattering corresponding to disk temperatures around $10^6-10^7$ K, the non-zero torque near the inner disk edge associated with strong accretion flow magnetization, and returning radiation effects. Fitting the Cygnus X-1 observational data using the radial temperature of Eq. (\ref{temp_A}) yields a color correction factor of $f_{col}\approx 1$ and a substantially lower mass accretion rate than what would be otherwise necessary to produce the disk photons above 1 keV. 

\begin{table*}
\caption{\label{Table2} Model parameterization of Cygnus X-1 that fits the observational data.}
\begin{center}
\begin{tabular}{l l l l}
\hline \\ [0.005ex] 
\multicolumn{4}{c}{Accretion disk-Corona}\\ 
\hline \\ [0.01ex]
Parameter & High/Soft & Hard/Low & Description \\
\hline \\ [0.01ex]
$R_{in}/R_{ISCO}$ & 1 & 20 & inner disk boundary \\ [0.2ex]
$\dot{m}$ & 0.007 & 0.004 & mass accretion rate \\ [0.2ex]
$R_{cor}/R_g$ & 20 & 60 & corona radius \\ [0.2ex]
$T_{cor}$ [keV] & 50 & 150 & corona temperature \\ [0.2ex]
$\tau _{cor}$ & 0.29 & 0.50 & thermal corona optical depth \\ [0.2ex]
p & 4.9 & 1.8 & corona power-law index \\ [0.2ex]
$\gamma _{min}$ & 1.8 & 1.7 & minimum energy of non-thermal electrons \\ [0.2ex]
$\gamma _{max}$ & >20 & >26 & maximum energy of non-thermal electrons \\ [0.2ex]
h & 0.050 & 0.294 & non-thermal/thermal electrons ratio in the corona \\ [0.2ex]   
\hline \\ [0.005ex] 
\multicolumn{4}{c}{Relativistic Jet}\\ 
\hline \\ [0.01ex]
$z_i$ [cm] & & $3\times 10^8$ & base of the jet \\ [0.2ex]
$z_f/z_i$ & & 3 & acceleration zone length \\ [0.2ex] 
$q_{rel}$ & & $3\times 10^{-5}$ & ratio of accelerated to thermal particles \\ [0.2ex]
$\eta _{acc}$ & & 0.15 & acceleration efficiency \\ [0.2ex]
$L_p/L_e$ & & 0.009 & ratio of the energy carried by hadrons to leptons \\ [0.2ex]
$\beta_{jet}$ & & 0.92 & bulk velocity \\ [0.2ex]
$\xi$ [$^{\circ}$] & & 1.2 & half-opening angle \\ [0.2ex]
\hline
\end{tabular}
\end{center}
\end{table*} 

Moreover, the hard tail above 5 keV is reproduced by the inverse-Compton (IC) scattered disk emission off a thermal/non-thermal corona hybrid described by the parameters of Table \ref{Table2}. In this case, the corona is much more weakly present compared to the hard/low state. The power-law spectrum indicates an electron (or positron) distribution injection, differentiating the two leading emission mechanisms near 10 keV. The injected distribution could be a product of an intense photon annihilation process above the disk level that softens by gradually losing energy due to IC scattering. The resulting power-law has an index of $p>3$ since $p_{inj}>2$.   We neglect Compton reflection of the scattered spectrum and irradiation effects beyond those integrated into the \texttt{kerrbb} model as they overcome the purpose of this work. As a result, there is a slight deviation near 6-7 keV in Fig. \ref{figure9} where the iron $K$ line is detected \citep{Walton_2016}. 

We assume the injection of 50 non-thermal electrons blending in a thousand thermal ones. Those particles preserve their initial power-law shape, while the rest are described by a Maxwell–Jüttner distribution, which dominates for energies below $\gamma =1.8$. The optical depth in Table \ref{Table2} corresponds to the thermal counterpart of the corona hybrid. Assuming a fixed spherical region, injecting power-law particles increases corona's opacity by a factor of $1+h$, where $h$ is the fraction of accelerated to thermal electrons (or positrons) (i.e., $\tau _{cor}^h= \tau _{cor}(1+h)\approx 0.3$ in our case).        

\subsection{Jet emission model}

We adopt a simple jet model that assumes partial acceleration of hadrons and leptons due to shock-wave propagation within a conic slice of the jet frame. A power-law energy dependence with an index of $p=2$ describes the non-thermal particle distribution \citep{Bosch-Ramon2006, REYNOSO2019}. The model is explicitly described in previous works \citep{Reynoso, Romero2007, Papavasileiou2021, Papavasileiou2022, Kosmas_2023}. The transport equation shaping the energy-dependent particle distributions within the jet is in a steady-state form as
\begin{equation}
\frac{\partial N(E,z)b(E,z)}{\partial E}+t_{dec}^{-1}N(E,z)=Q_{inj}(E,z) .
\label{tranf-equat}
\end{equation}
The particle injection function is denoted as $Q_{inj}(E,z)$ and the rate at which the particles lose energy is given by $b(E)=-Et_{loss}^{-1}$.     

The accelerated electrons and protons inside the jet are cooled down primarily via synchrotron emission, which covers the regime from radio to soft gamma-ray wavelengths. On the other hand, the hadronic emission mechanism involves collisions between the relativistic protons and thermal ones, which produce pions and muons that decay to high-energy neutrinos and gamma-rays.  

As described in the previous subsection, we consider an IC spectrum component to explain the MeV tail observed from Cygnus X-1. We employ the mathematical formalism described in \citet{Rybicki_2008} for those calculations, assuming both the Compton and inverse Compton cases. We adopt the conclusions of \citet{Blumenthal} to estimate the scattered spectrum caused by an electron distribution with $\gamma _e\gg 1$ while considering both the Thomson and the Klein-Nishina limits.

\subsection{Cygnus X-1: hard/low spectral state}

The hard/low state is characterized by the intense presence of the relativistic jet and a strong coronal activity. The accretion disk experiences a truncation to larger radii, leaving behind a radiatively inefficient advection-dominated accretion flow (i.e., ADAF) \citep{Takahashi_2007, Narayan_2008, Li_2023}. Therefore, relativistic effects are expected to be less significant compared to the soft state.

\begin{figure}[ht] 
\begin{tikzpicture}
\centering
  \node (img1)  {\includegraphics[width=\linewidth]{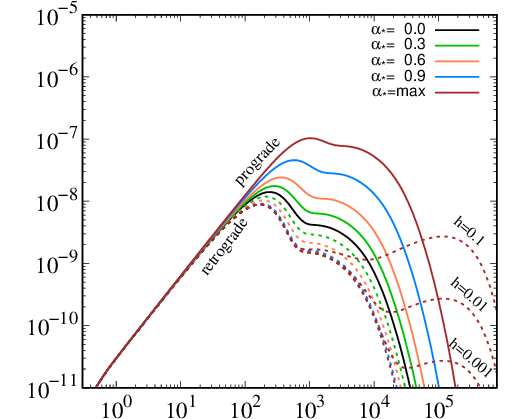}};
  \node[left= of img1, node distance=0cm, rotate=90, anchor=center,yshift=-1.4cm, font=\color{black}, font=\normalsize] {$\epsilon F_{\epsilon}$ [$ergs^{-1}cm^{-2}$]};
  \node[below= of img1, node distance=0cm, yshift=1cm, xshift=0.5cm, font=\color{black}, font=\normalsize] {$E$ [$eV$]};
\end{tikzpicture}

\caption{\label{figure10} The thermal disk emission based on the dimensionless spin parameter. The spectrum produced when seed photons from the disk are IC scattered off the purely thermal corona is also included. The dotted brown curves correspond to the injection of non-thermal particles in the corona, assuming a maximum counter-rotating black hole. Their number density is written as a fraction $h$ of the thermal electrons.}
\end{figure}

In Fig. \ref{figure10}, we present the total disk spectra for various spin values of the black hole. They comprise of a black-body component and the respective IC emission due to a purely thermal corona. For the case of a rapidly counter-rotating black hole, we also include the scattered emission flux due to a non-thermal corona component described by the parameter $h$, which is the number density ratio of the non-thermal to thermal particles. The corona parameterization is listed in Table \ref{Table2}. In this case, we assume an accretion rate of $\dot{M}\approx 10^{-8}$ \(M_\odot\)/yr. 

The slump between the two emission peaks in Fig. \ref{figure10} is smoothed out for rapidly rotating black holes. At the same time, the incident disk spectrum covers a broader energy range, allowing for up-scattering even to the gamma-ray regime. Nonetheless, this outcome heavily relies on the assumed truncation radius, which we consider to be equal to $3R_{ISCO}$ in this case. Thus, the hard state signifies the growing irrelevance of the relativistic effects on the detected emission flows due to the disk retreating to distances further away from the black hole.     

\begin{figure}[ht] 
\begin{tikzpicture}
\centering
  \node (img1)  {\includegraphics[width=\linewidth]{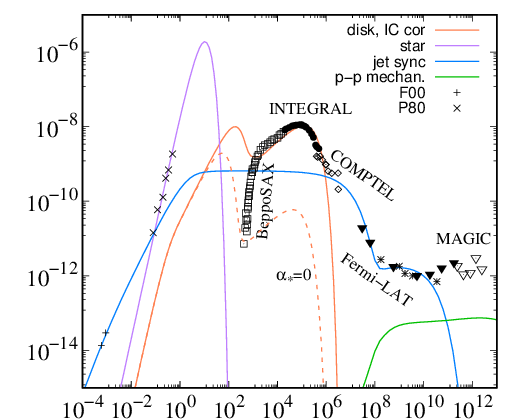}};
  \node[left= of img1, node distance=0cm, rotate=90, anchor=center,yshift=-1.4cm, font=\color{black}, font=\normalsize] {$\epsilon F_{\epsilon}$ [$ergs^{-1}cm^{-2}$]};
  \node[below= of img1, node distance=0cm, yshift=1cm, xshift=0.5cm, font=\color{black}, font=\normalsize] {$E$ [$eV$]};
\end{tikzpicture}

\caption{\label{figure12} The spectral energy distribution associated with the hard/low state of Cygnus X-1. The plotted observational data include the IR and radio emission points up to 1 eV (\citet{Persi_1980} as P80, \citet{Fender_2000} as F00), the soft and hard X-ray emission detected by BeppoSAX and INTEGRAL \citep{DiSalvo_2001, Zdziarski_2012}, and the gamma-ray measurements and upper limits by COMPTEL, Fermi-LAT, and MAGIC \citep{McConnell_2002, Malyshev_2013, Ahnen2017}.}
\end{figure}

The IR measurements of Cygnus X-1 \citep{Persi_1980}, as demonstrated in Fig. \ref{figure12}, are fitted by the thermal emission of the system's blue supergiant, whose effective temperature and luminosity are listed in Table \ref{Table1}. Synchrotron emission by magnetically-accelerated leptons and hadrons inside the jet frame reproduce the radio emission flux below 0.01 eV. Nevertheless, the employed jet emission model is relatively simple and would require some necessary modifications to replicate the softer radio data accurately.

The soft and hard X-ray measurements by BeppoSAX and INTEGRAL \citep{DiSalvo_2001, Zdziarski_2012} are fitted by an IC spectrum caused by the corona's intervention on the incident disk spectrum. The corona parameterization for the hard state is given in Table \ref{Table2}. It is significantly hotter, larger, and optically thicker than the one attributed to the soft state. Moreover, the thermal electrons dominate the distribution over energies up to $\gamma =1.7$. 

So far, we have adopted a maximally rotating compact object. Matching the observations with a static black hole (i.e., $\alpha _*=0$) would require a considerably higher accretion rate and the adoption of a color-corrected disk. In addition, the corona electrons would have to achieve energies above $\gamma =40$, as indicated in Fig. \ref{figure12}.     

The intersection between the high-energy tail of the synchrotron spectrum and the IC scattered flux matches the MeV measurements by COMPTEL \citep{McConnell_2002}. Without the particle injection in the corona, the thermal distribution would have to reach a temperature of a few hundred keV to replicate the data. 

The synchrotron component due to highly relativistic electrons and protons (the latter achieve energies up to $E_{max}\approx 10^7$ GeV) confirms the Fermi-LAT gamma-ray measurements and upper limits, as seen in Fig. \ref{figure12}. At last, the proton-proton mechanism is too weak in our lepton-dominated jet model (i.e., $L_p/L_e=0.009$) to reach the upper limits set by the MAGIC collaboration for the high-energy gamma-ray regime \citep{Ahnen2017}. Nevertheless, photo-absorption caused by the disk is dominant near the lower-jet emitting regions. Hence, detected gamma-ray fluxes above 10 GeV are more likely to originate inside the expanded jet frame \citep{Zanin2016, Papavasileiou_AA}.           

\section{Summary and conclusions}\label{Sec5}

We propose a simple and effective approach to include relativistic effects in the broad emission from standard accretion disks. That is possible by integrating the emission pattern of the \texttt{kerrbb} model, found in \citet{Campitiello_2018}, to the observed spectra via modification of the disk's boundary, through which most of the gravitational energy is released. Hence, it is necessary to implement spin-dependent temperature profiles derived from pseudo-Newtonian potentials that account for the Kerr metric’s fundamental properties. Furthermore, we employ this approach to fit the observed spectra from Cygnus X-1 while going through its i) high/soft and ii) hard/low spectral states.

The energy distribution radiated away from a thin accretion disk described by the Novikov-Thorne and  \texttt{kerrbb} models is equivalent to the one emitted by a Shakura-Sunyaev disk with a modified innermost radius, which depends on the system's spin and viewing angle. It is crucial to use the spin-dependent PN potential. Otherwise, the modified boundary cannot be applied in XRBs inclined by $\theta >60^{\circ}$, independently of the black hole's rotation. The best-suited potential is the one suggested by \citet{Artemova_1996} because it produces valid results for small radii, even more so than the $R_{ISCO}$. Hence, the boundary transition, unconstrained by the spin parameter, is within the $R_{in}/R_{ISCO}= 0.2-1.0$ range, when $\theta \geq 60^{\circ}$, and $R_{in}/R_{ISCO}= 1.0-2.0$, otherwise. 

The thermal emission from Cygnus X-1 in the high/soft state is reproduced by a black-body spectrum associated with the inner disk surface. The non-thermal tail at higher energies is fitted by the IC scattered emission of the incident disk photons due to a thermal/non-thermal corona hybrid. Regarding the hard/low state, the synchrotron emission from a fairly simple lepto-hadronic jet is sufficient to match the radio and soft gamma-ray measurements. The soft X-ray emission approximates the black-body spectrum from a truncated accretion disk. It is accompanied by a harder IC component due to a more extensive, hotter, and optically thicker corona compared to the one in the soft state. Regarding higher energies, the collisions of protons inside the jet are incapable of generating detectable high-energy gamma-rays.    

Our results show a very good agreement with the fully relativistic treatment of the Novikov-Thorne model incorporated in the \texttt{kerrbb} code, with a deviation within 0.03-4\%. The intrinsic error of the pseudo-potential, which leads to an overestimation of the radiated luminosity, is minimized by the consistent boundary transition to fit the relativistic predictions. However, the inaccuracy in the radiative efficiency, assuming an inner radius of $R_{ISCO}$, remains. Therefore, we propose a fixing factor of $k\approx 1.075$ that marginally shifts the $R_{ISCO}$ to match the theoretical predictions.

\bibliography{references}
\begin{appendix}
\section{Derivation of the temperature corresponding to each disk surface area}
\label{Temp_deriv}
The temperature corresponding to the black-body spectrum emitted by a single surface area of the accretion disk, according to the Stefan-Boltzmann law, is given by   
\begin{equation}
T(R)= \left(\frac{Q(R)}{\sigma}\right)^{1/4} ,
\end{equation}
where $\sigma $ is the well-known Stefan-Boltzmann constant and $Q(R)$ is the energy flux generated per unit of time and area of the disk side that faces the Earth. The latter is defined as
\begin{equation}
Q(R)= \frac{-\dot{M}(j-j_{in})}{4\pi R}\frac{d\omega}{dR} , 
\end{equation}
where $j_{in}$ is the specific angular momentum at the innermost disk regions and $\omega$ is the angular velocity, which is determined as 
\begin{equation}
\omega = \left(\frac{1}{R}\frac{d\Phi}{dR}\right)^{1/2} .
\end{equation}
Replacing the angular velocity and momentum per unit mass yields the following definition of the disk's surface brightness
\begin{equation}
Q(R)= -\frac{\dot{M}}{8\pi}\left(\frac{dF}{dR}-\frac{F}{R}\right)\left(1-\sqrt{\frac{F(R_{in})R_{in}^3}{F(R)R^3}}\right) ,
\end{equation}
where $F(R)$ is the gravitational force.

\section{Radiative efficiency configuration due to general relativistic effects}
\label{f_func}

The analytic function integrating all the effects of general relativity is presented in \citet{Campitiello_2018} as
\begin{equation}
f(\theta , \alpha _*)= A\cos\theta\left[1-(\sin\theta )^C\right]^B\left[1-E(\sin\theta )^F\right]^D .
\end{equation}
The parameters in the above expression depend on the spin. Posing $\xi =\log (1-\alpha _*)$ yields the following best-fitting functions for the numerical results of \texttt{kerrbb}
\begin{align*}
&\log A= 0.21595 +0.09107\xi -0.05037\xi ^2-0.02739\xi^3-0.00361\xi^4 , \\
&B= -0.20229 +0.17538\xi -0.14390\xi ^2-0.14534\xi^3-0.04544\xi^4-0.00480\xi^5 , \\
&C= 1.92161 +0.27712\xi +0.67368\xi ^2+0.81327\xi^3+0.48946\xi^4+0.13591\xi^5+0.01373\xi^6 , \\
&D= 0.12120 -0.07852\xi +0.08995\xi ^2+0.12746\xi^3+0.04556\xi^4+0.00510\xi^5 , \\
&E= 0.95973 -0.02003\xi +0.09341\xi ^2+0.16775\xi^3+0.11440\xi^4+0.03367\xi^5+0.00351\xi^6 , \\
&F= 6.62190 -3.84845\xi -3.11662\xi ^2-3.61394\xi^3-1.54083\xi^4-0.19834\xi^5 .
\end{align*}

\end{appendix}

\end{document}